\documentclass[pre,preprint,showpacs]{revtex4} \usepackage{graphicx}
\usepackage{graphicx}
\begin{document}
\title{Classical analysis of phase-locking transients and Rabi-type
oscillations in microwave-driven Josephson junctions}
\author{Jeffrey E.~Marchese}
\affiliation{Department of Applied Science, University of California,
Davis, California 95616}
\author{Matteo Cirillo}
\affiliation{Dipartimento di Fisica and INFM, Universit\`{a} di Roma
"Tor Vergata", I-00173 Roma, Italy}
\author{Niels Gr{\o}nbech-Jensen}
\affiliation{Department of Applied Science,$\;$University of California,
Davis,  California 95616}
\date{\today}
\begin{abstract}
We present a classical analysis of the transient response of Josephson
junctions perturbed by microwaves and thermal fluctuations. The results
include a specific low frequency modulation in phase and amplitude behavior
of a junction in its zero-voltage state. This transient modulation frequency
is linked directly to an observed variation in the probability for
the system to switch to its non-zero voltage state.
Complementing previous work on linking classical analysis to the
experimental observations of Rabi-oscillations, this expanded perturbation
method also provides closed form analytical results for attenuation of
the modulations and the Rabi-type oscillation frequency.
Results of perturbation analysis are compared directly (and quantitatively)
to numerical simulations of the classical model as well as published
experimental data, suggesting that transients to phase-locking are
closely related to the observed oscillations.
\end{abstract}
\maketitle
\section{Introduction}
Much experimental attention has recently been given to the topic
of Rabi-oscillations \cite{Rabi} in microwave irradiated Josephson
systems at very low temperatures
\cite{Martinis02,Berkley03,Claudon04,Simmonds04}.
A common motivation for the studies is the proposed macroscopic
quantum behavior outlined in Ref.~\cite{Caldeira81}, which has opened
the way for potential applications
of Josephson technology in quantum information processing \cite{Legget04}.
This is proposed by operating a Josephson junction in its zero-voltage state at
temperatures below the quantum transition temperature, while
manipulating the possible energy states with the application of commensurate
microwave frequencies. The significance of Rabi-oscillations is many fold,
and includes {\it i}) a direct connection to a well known quantum mechanical
concept in perturbed atomic systems, and {\it ii}) a method by which the control
of quantum states of a system can be evaluated. Thus, it is of significant
importance to understand the nature of these observations in order for
us to evaluate how to interpret the Josephson system under investigation.

Recent classical analyses of Josephson junctions, perturbed by microwaves
and low temperature thermal fluctuations, have revealed that
key signatures of microwave-induced multi-peaked switching distributions
\cite{Clarke85,Friedman00,Wallraff03_1,Wallraff03_2},
used to display quantum
mechanical features of Josephson junctions, have direct classical
analogs, which may obscure the interpretation
of the observations \cite{Jensen04_1,Jensen04_2,Jensen04_3,Cirillo05}.
Inspired by the
work on slowly modulated transients to phase-locking in Josephson systems
by Lomdahl and Samuelsen \cite{Lomdahl88}, it has most recently
been demonstrated \cite{Jensen05} that also Rabi-oscillations
have a classical analogue in microwave perturbed Josephson junctions,
providing a system response very similar to reported observations
under the same conditions. We therefore denote this classical phenomenon
Rabi-type oscillations. It is the aim of this
paper to complement and expand on the analysis provided in Ref.~\cite{Jensen05}
in order to extract more detailed information from the classical model
as it relates to observations of Rabi-oscillations. We further demonstrate
the validity of the analysis through direct quantitative comparisons with
the observations of Ref.~\cite{Claudon04}, in which enough information
is provided to conduct direct quantitative comparisons with our theory.
Specifically, we present
an analysis based on direct perturbations in the dynamical variables. This
approach results in closed form results that include dissipaion, which
is of direct importance for experimental measurements of, e.g, coherence
time.

We will start section II by providing the system equations that govern
the classical model as it relates to specific reported measurements of
Rabi-oscillations. Next, in section III, we present a new, more complete,
perturbation
analysis of the slow transient modulations to phase-locking, and provide
specific analytical results for (Rabi) oscillation frequency and
attenuation as a function of the system parameters. The theory is numerically
validated in section IV. Direct observations of Rabi-type oscillations
in the classical model are produced in section V through numerical
simulations of a system parameterized to mimic reported experiments,
and close agreement is found between these simulations, the developed theory,
and the reported experiments. We conclude the paper in section VI.

\section{Classical Model}
A normalized classical equation for a Josephson junction can be written
\cite{Barone82}
\begin{eqnarray}
\ddot{\varphi}+\alpha\dot{\varphi}+\sin{\varphi} & = &
\eta+\varepsilon_s(t)\sin(\omega_st+\theta_s)+\varepsilon_p(t)+n(t) \; ,
\label{eq:Eq_1}
\end{eqnarray}
where $\varphi$ is the difference between the phases of the quantum mechanical
wave functions defining the junction, $\eta$ represents the dc bias current,
and $\varepsilon_s(t)$ and $\omega_s$ represent microwave current amplitude
and frequency, respectively. All currents are normalized to the
critical current $I_c$, and time is measured in units of the inverse
plasma frequency $\omega_0^{-1}$, where
$\omega_0^2=2eI_c/\hbar C=2\pi I_c/\Phi_0C$,
$C$ being the capacitance of the junction and $\Phi_0=h/2e$ the flux quantum.
The phase of the applied microwave field is $\theta_s$. Tunneling of
quasiparticles
is represented by the dissipative term, where $\alpha=\hbar\omega_0/2eRI_c$
is given by the shunt resistance $R$, and the accompanying thermal
fluctuations are defined by the dissipation-fluctuation relationship
\cite{Parisi88}
\begin{eqnarray}
\langle n(t) \rangle & = & 0 \label{eq:Eq_2} \\
\langle n(t)n(t^\prime) \rangle & = & 2\alpha\frac{k_BT}{H_J}\delta(t-t^\prime)
\; , \label{eq:Eq_3}
\end{eqnarray}
$T$ being the temperature and $H_J=I_c\Phi_0/2\pi$ is the characteristic
Josephson energy. A current pulse for probing the state of the system
is represented by $\varepsilon_p(t)$.

\begin{figure}[h]
\begin{center}
\scalebox{.18}{\centering \includegraphics{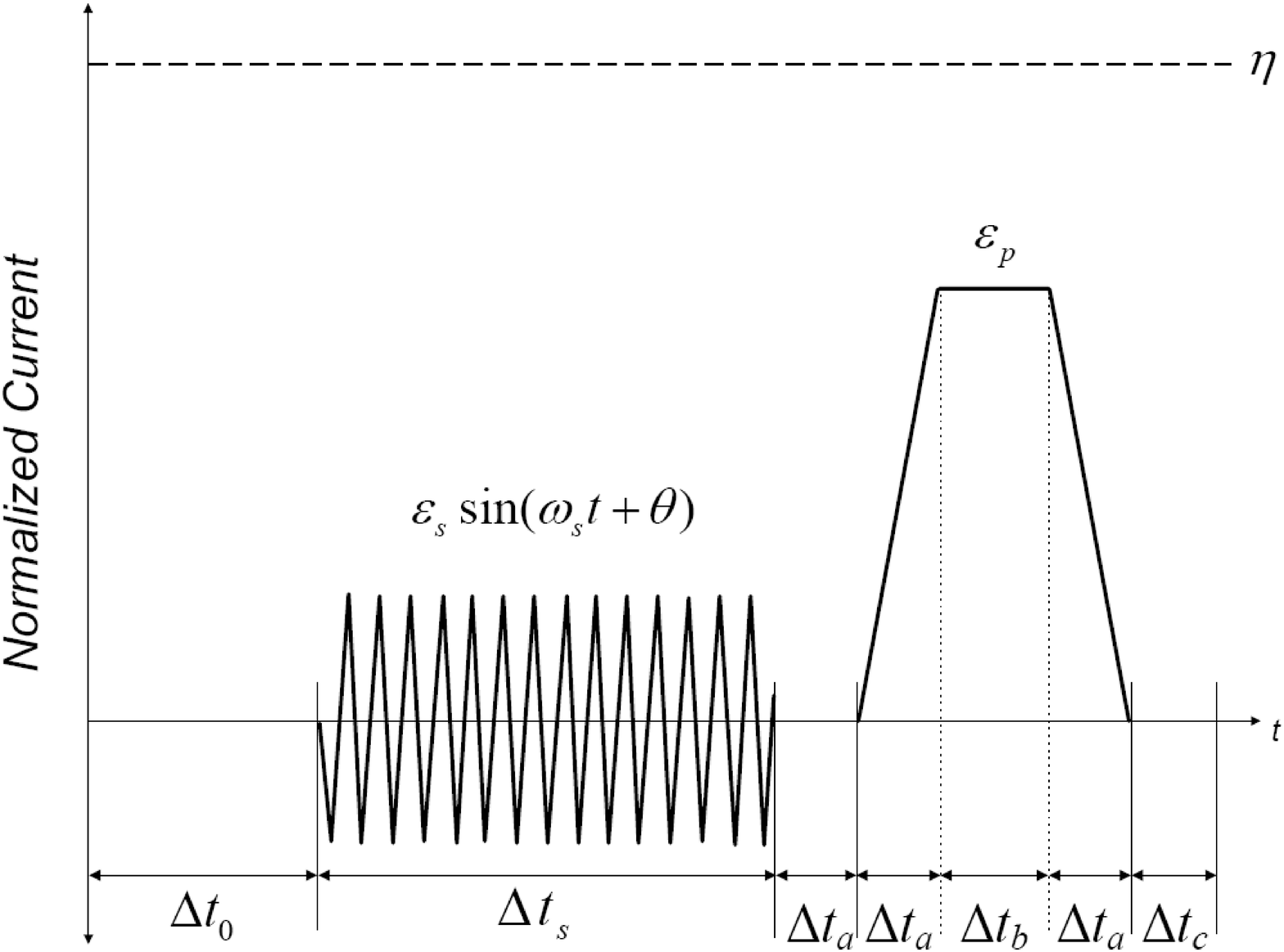}}
\end{center}
\vspace{-0.0cm}
\caption{Schematic of signaling protocol, which is adopted from
Ref.~\cite{Claudon04}. Dotted line represents the dc 
bias $\eta$ which is constant throughout each run. The normalized
time ineterval $\Delta t_0$ represents a 91 ns
duration during which no ac signal is present to allow the junction to
achieve a steady-state. $\Delta t_s$ represents the duration of the ac signal.
This value is varied so as to observe the temporal Rabi-type oscillations.
Values for the signal timings are $\Delta t_a=1.5ns\omega_0$,
$\Delta t_b=2.5ns\omega_0$, and $\Delta t_c=2.0ns\omega_0$.
}
\label{fig:fig1}
\end{figure}

From the above normalizations, we define the normalized energy from the
time independent perturbation terms in Equation~(\ref{eq:Eq_1})
\begin{eqnarray}
H & = & \frac{1}{2}\dot{\varphi}^2+1-\cos{\varphi}-\eta\varphi \; .
\label{eq:Eq_4}
\end{eqnarray}
The undamped bias current dependent plasma frequency is given by
\begin{eqnarray}
\omega_J & = & \sqrt[4]{1-\eta^2} \; . \label{eq:Eq_6}
\end{eqnarray}
Within this model, we will make analysis and conduct simulations
according to experiments reported in Ref.~\cite{Claudon04}. System parameters,
such as $\eta$, $\alpha$, $\omega_s$, $\varepsilon_s(t)$, and $\varepsilon_p(t)$
are matched as closely as possible to reported values, and Rabi-type
oscillations are observed statistically through simulations of the 
distribution of probe field induced switching from the zero-voltage state
as a function of applied microwave field
$\varepsilon_s(t)\sin(\omega_st+\theta_s)$. Microwave and probe perturbations
have the form sketched in Figure~\ref{fig:fig1}.
We assume the phase $\theta_s$ of the
microwave field is random for each switching event. Notice the difference
in probe field between this presentation (along with Ref.~\cite{Claudon04})
compared to Refs.~\cite{Martinis02,Simmonds04,Jensen05}, where the probe was a
microwave field with frequency slightly lower than $\omega_s$.

\begin{figure}[h]
\begin{center}
\scalebox{.42}{\centering \includegraphics{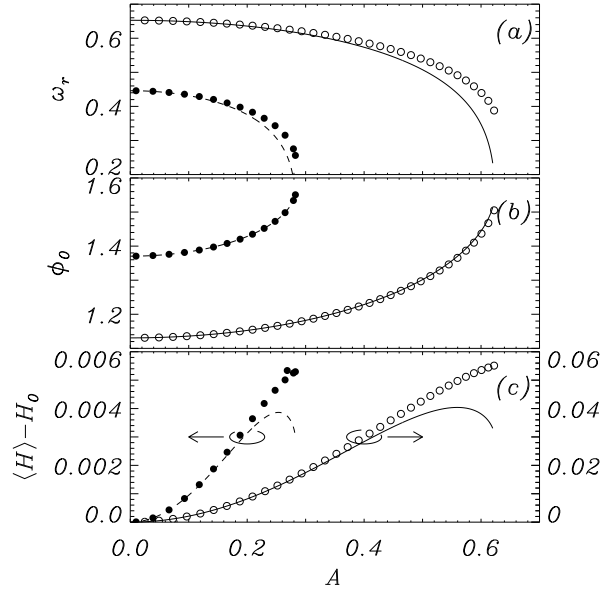}}
\end{center}
\vspace{-1.5cm}
\caption{Comparison between numerical results of Equation~(\ref{eq:Eq_1})
(markers) and the monochromatic ansatz Equation~(\ref{eq:Eq_7})
(curves) for $\alpha=\varepsilon_s(t)=\varepsilon_p(t)=T=0$.
Results for $\eta=0.904607$ are displayed with $\circ$ and solid curves, while
results for $\eta=0.98$ are displayed with $\bullet$ and dashed curves.
(a) Anharmonic relationship between amplitude $A$ and frequency $\omega_r$
Equation~(\ref{eq:Eq_9}); (b) amplitude $A$ and average phase $\varphi_0$
Equation~(\ref{eq:Eq_8}); and (c) amplitude $A$ and normalized energy $H-H_0$
Equations~(\ref{eq:Eq_28}) and (\ref{eq:Eq_29}).
}
\label{fig:fig2}
\end{figure}

\section{Perturbation Analysis}
Following the analysis of Refs.~\cite{Jensen04_2,Jensen04_3,Pedersen73},
we will use the
monochromatic ansatz for a solution of the unperturbed Equation~(\ref{eq:Eq_1})
($\alpha=\varepsilon_s(t)=T=\varepsilon_p(t)=0$)
\begin{eqnarray}
\varphi & = & \varphi_0+\psi = \varphi_0+A\sin(\omega t+\theta) \; ,
\label{eq:Eq_7}
\end{eqnarray}
where $\varphi_0$ is the average phase, $A$ is an oscillation amplitude (which
is not necessarily small), and
$\omega$ is the natural frequency. Inserting this ansatz into
Equation~(\ref{eq:Eq_1}), we obtain the anharmonic relationships between the
parameters
\begin{eqnarray}
&& J_0(A)\sin\varphi_0  =  \eta \label{eq:Eq_8} \\
&& \omega^2 = \omega_r^2  =  \frac{2J_1(A)}{A}
\sqrt{1-\left(\frac{\eta}{J_0(A)}\right)^2} \; , \label{eq:Eq_9}
\end{eqnarray}
where $J_n$ is the $n$th order Bessel function of the first kind \cite{GR}.
Notice that the ansatz dictates $J_0(A)>\eta$ as a condition for existence
of a zero-voltage state.
Also notice that $\omega_r\le\omega_J$ and that $\omega_r\rightarrow\omega_J$
for $A\rightarrow0$.
The average system energy of the monochromatic ansatz is found by inserting
Equations~(\ref{eq:Eq_7}) and (\ref{eq:Eq_8}) into (\ref{eq:Eq_4}), then
average over the period given by (\ref{eq:Eq_9}),
\begin{eqnarray}
\langle H\rangle_{\frac{2\pi}{\omega}} & = &
\frac{1}{4}A^2\omega^2+1-J_0(A)\cos\varphi_0-\eta\varphi_0 \; ,
\label{eq:Eq_28}
\end{eqnarray}
where the minimum energy $H_0$ for a given $\eta$ is
\begin{eqnarray}
H_0 & = & 1-\sqrt{1-\eta^2}-\eta\sin^{-1}\!\eta \;. \label{eq:Eq_29}
\end{eqnarray}
The applicability of this very simple monochromatic
ansatz (\ref{eq:Eq_7}) is illustrated in Figure~\ref{fig:fig2}. Here
we have shown the average phase $\varphi_0$ and oscillation frequency
$\omega_r$ as a function of oscillation amplitude $A$ for two different
values of dc bias $\eta$. Data are obtained from direct numerical simulations
of Equation~(\ref{eq:Eq_1}) (markers) and from
Equations~(\ref{eq:Eq_8})-(\ref{eq:Eq_29}) (curves).
The agreement between the results of the ansatz and the simulations is
very close, except at the highest amplitudes (when $J_0(A)\rightarrow\eta$),
where higher harmonics of
the dynamics contribute significant components to the dynamics.
We use this ansatz to develop a perturbation
theory for the system response to the application of microwaves and
damping.

For $T=\varepsilon_p(t)=0$ we now separate the parameters of the
ansatz into steady state and small time dependent deviations;
$A=\bar{A}+\delta{\!A}$, $\theta=\bar{\theta}+\delta{\theta}$,
and $\varphi=\bar{\varphi}_0+\delta\varphi_0$, where $|\delta{\!A}|\ll1$,
$|\delta{\theta}|\ll1$, and $|\delta\varphi_0|\ll1$.
Inserting Equation~(\ref{eq:Eq_7}) into (\ref{eq:Eq_1}) for
$\varepsilon_p(t)=T=0$
and $\varepsilon_s(t)=\Theta(t)\varepsilon_s$ ($\Theta$ being Heavyside's
step function), yields the steady state phase-locked relationships
\cite{Jensen04_2,Jensen05}
\begin{eqnarray}
\varepsilon_s^2 & = & \bar{A}^2\left((\omega_s^2-\omega_r^2)^2+
\alpha^2\omega_s^2\right) \label{eq:Eq_10} \\
\tan(\bar{\theta}-\theta_s) & = &
\frac{\alpha\omega_s}{\omega_s^2-\omega_r^2} \; , \label{eq:Eq_11}
\end{eqnarray}
and the linearized expressions for the small deviations from steady state:
\begin{eqnarray}
&&\ddot{\delta\varphi}_0+\alpha\dot{\delta\varphi}_0+
J_0(\bar{A})\cos\bar{\varphi}_0\,\delta\varphi_0 =
J_1(\bar{A})\sin\bar{\varphi}_0\,\delta{\!A} \label{eq:Eq_12} \\
&&\ddot{\delta{\!A}}+\alpha(\dot{\delta{\!A}}-\omega_s\bar{A}\delta\theta)-
2\omega_s\bar{A}\dot{\delta\theta}+\left[\left(J_0(\bar{A})-J_2(\bar{A})\right)
\cos\bar{\varphi}_0-\omega_s^2\right]\,\delta{\!A} =
2J_1(\bar{A})\sin\bar{\varphi}_0\,\delta\varphi_0 \nonumber \\
&& \label{eq:Eq_13} \\
&&\bar{A}\ddot{\delta{\theta}}+\alpha(\bar{A}\dot{\delta{\theta}}+
\omega_s\delta{\!A})+2\omega_s\dot{\delta{\!A}}+\left[\left(J_0(\bar{A})+
J_2(\bar{A}\right)\cos\bar{\varphi}_0-\omega_s^2\right]\bar{A}\,\delta{\theta} =
0 \; . \label{eq:Eq_14}
\end{eqnarray}
Equation~(\ref{eq:Eq_12}) represents the small amplitude,
slowly varying terms in
(\ref{eq:Eq_1}), while (\ref{eq:Eq_13}) and (\ref{eq:Eq_14}) represent
small amplitude terms oscillating with $\omega_s$. Evaluating a solution
to Equations~(\ref{eq:Eq_12})-(\ref{eq:Eq_14}) is not simple, but making the
assumption that all three variables oscillate slowly with the same frequency
$\Omega_R$ and decay with the
same attenuation $\beta$, we can write the simple relationships:
$\delta{\!A}=\exp(i\Omega t)$, $\bar{A}\delta{\theta}=\kappa\delta{\!A}$,
and $\delta\varphi_0=\gamma\delta{\!A}$, with $\Omega=\Omega_R+i\beta$.
This results in the following polynomial
\begin{eqnarray}
&& \Omega^6 - 3i\alpha\Omega^5-\left[\frac{15}{4}\alpha^2-a_4\right]\Omega^4
+i\alpha\left[\frac{5}{2}\alpha^2-2a_4\right]\Omega^3 
+ \left[\frac{15}{16}\alpha^4-\frac{3}{2}\alpha^2a_4+a_2\right]\Omega^2
\nonumber \\
&&-i\alpha\left[\omega_s^2\alpha^2+\Gamma_1\right]\Omega
= \alpha^2\omega_s^2J_0(\bar{A})\cos\bar{\varphi}_0+\Gamma_2 \; ,
\label{eq:Eq_15}
\end{eqnarray}
where the parameters $a_i$ and $\Gamma_i$ are given in 
Equations~(\ref{eq:Eq_17})-(\ref{eq:Eq_21}) below.
The polynomial can be simplified considerably by inserting
$\Omega=\Omega_R+i\beta$, and realizing that $\beta=\frac{1}{2}\alpha$
is the complex part of the roots (for all roots) with a non-zero real part
$\Omega_R$; i.e., for underdamped modulations. Notice that dissipative
systems ($\alpha>0$) will naturally imply that
$\Omega_R^2\rightarrow0_+$ for a nonzero threshold value of the microwave
amplitude $\varepsilon_s>0$ \cite{Lomdahl88,Jensen05}.
With $\beta=\frac{1}{2}\alpha$ the (real) modulation frequency $\Omega_R$
is determined by the following real polynomial:
\begin{eqnarray}
\Omega_R^6+a_4\Omega_R^4+a_2\Omega_R^2+a_0 & = & 0 \; , \label{eq:Eq_16}
\end{eqnarray}
where
\begin{eqnarray}
a_4 & = & \frac{3}{4}\alpha^2-2\omega_s^2-3J_0(\bar{A})\cos\bar\varphi_0
\label{eq:Eq_17} \\
a_2 & = & \frac{3}{16}\alpha^4-
\frac{3}{2}\alpha^2J_0(\bar{A})\cos\bar{\varphi}_0+\Gamma_1
\label{eq:Eq_18} \\
a_0 & = & \frac{1}{64}\alpha^6-\frac{1}{16}\alpha^4a_4+\alpha^2(\frac{a_2}{4}-
\omega_s^2J_0(\bar{A})\cos\bar{\varphi}_0)-\Gamma_2 
\label{eq:Eq_19} \\
\Gamma_1 & = & (\omega_s^2-\bar\omega_r^2)
\left(\omega_s^2+\bar\omega_r^2-2J_0(\bar{A})\cos\bar\varphi_0\right)
+ 2(J_0(\bar{A})\cos\bar\varphi_0+\omega_s^2)J_0(\bar{A})\cos\bar\varphi_0
\nonumber \\
&& -2J_1^2(\bar{A})\sin^2\!\bar\varphi_0 \label{eq:Eq_20} \\
\Gamma_2 & = & (\omega_s^2-\bar\omega_r^2)
\left[\left(\omega_s^2+\bar\omega_r^2-
2J_0(\bar{A})\cos\bar\varphi_0\right)J_0(\bar{A})\cos\bar\varphi_0+
2J_1^2(\bar{A})\sin^2\!\bar\varphi_0\right] \; . \label{eq:Eq_21}
\end{eqnarray}
The steady-state resonance frequency $\bar\omega_r$ is given by
Equation~(\ref{eq:Eq_9}) as a function of $\bar{A}$.
Generally speaking, the three roots $\Omega_R^2$
are real and positive. One is located
near $(2\omega_s)^2$, one near $\omega_s^2$, and one much smaller than
$\omega_s^2$. It is the latter we are interested in when studying transients
and modulations to phase-locking, and it can be very well approximated by
either of the following approximations, since the normalized frequency is small:
\begin{eqnarray}
\Omega_R^2 & \approx & -\frac{a_0}{a_2} \label{eq:Eq_apprx1} \\
\Omega_R^2 & \approx & \frac{-a_2+\sqrt{a_2^2-4a_0a_4}}{2a_4}
\label{eq:Eq_apprx2}
\end{eqnarray}

A further
simplification to the coefficients in the polynomial can be produced by
noticing that the damping parameter $\alpha$ in relevant Josephson experiments
is very small, allowing for omission of several terms of high order in
$\alpha$ in Equations~(\ref{eq:Eq_17})-(\ref{eq:Eq_19}). A specific limit of
the above solution can be derived for small oscillation ($A$) and
microwave ($\varepsilon_s$) amplitudes for $\alpha=0$ and
$\omega_s^4\equiv1-\eta^2$
\begin{eqnarray}
\Omega_R & \approx & \sqrt{\frac{\Gamma_2}{\Gamma_1}} \label{eq:Eq_22} \\
& \rightarrow & \bar{A}^2\, \frac{\sqrt{3}}{16}\frac{2-\omega_s^4}{\omega_s^3}
=\bar{A}^2\, \frac{\sqrt{3}}{16}\frac{1+\eta^2}{(1-\eta^2)^{\frac{3}{4}}} \; , \; \; {\rm for} \; \; A\rightarrow0 \label{eq:Eq_23} \\
& = & \varepsilon_s^{\frac{2}{3}} \, \frac{\sqrt{3}}{4} \frac{(2-\omega_s^4)^{\frac{1}{3}}}
{\omega_s^{\frac{5}{3}}} = 
\varepsilon_s^{\frac{2}{3}} \, \frac{\sqrt{3}}{4} \frac{(1+\eta^2)^{\frac{1}{3}}}{(1-\eta^2)\omega_s^{\frac{5}{12}}} \; ,
\; \; {\rm for} \; \; \varepsilon_s\rightarrow0 \; . \label{eq:Eq_24}
\end{eqnarray}
The resonant choice $\omega_s^4\approx1-\eta^2$ is consistent with
the experimental studies \cite{Claudon04}. In fact, measurements are
usually conducted for $\omega_s$ very close to the linear resonance frequency
(see, e.g., Ref.~\cite{Martinis02,Simmonds04}). However, if
$\omega_s^4\neq1-\eta^2$, then $A\propto\varepsilon_s$, and the relationship
between $\Omega_R^2$ and $\varepsilon_s^2$ becomes linear for small values of
$A$ and $\varepsilon_s$. Further, if $\omega_s^4\lesssim1-\eta^2$, $A$
may become a multi-valued function of $\varepsilon_s$, corresponding to
several resonant energy states of the system for the same parameter values
\cite{Jensen05}.

The solution to Equation~(\ref{eq:Eq_16}),
which can be given in closed form as a solution to a third order polynomial,
is a complementary, more convenient, approach to the one presented
in Ref.~\cite{Jensen05}.
It provides the important additional information regarding the attenuation
$\beta$ of the transient modulations and it provides explicit expressions
for the modulation frequency as a function of all system parameters.
The two different perturbation methods agree
qualitatively, and are quantitatively similar. The main difference between
the two approaches is that we have here considered phase $\theta$ and amplitude
$A$ as harmonically varying variables, instead of using the energy balance
approach to transients \cite{Lomdahl88}, which treats the total energy of
the system as the linearized dynamical variable. Notice that the
linearized Equations~(\ref{eq:Eq_12})-(\ref{eq:Eq_14}) provide
information about frequency,
attenuation, and internal phase-relationships, but not about the specific
magnitude
of the deviations. However, the magnitude can be quantitatively estimated
from the following reasoning. For a simulation, in which a Josephson junction
is described by Equation~(\ref{eq:Eq_1}) with $\varepsilon_p(t)=T=0$ (or at
least very low temperature) and $\varepsilon_s(t)=\Theta(t)\varepsilon_s$,
the system
is at rest $\varphi=\varphi_0=\sin^{-1}\eta$ for $t<0$ ($A=0$). The onset
of the microwave field at $t=0$ will therefore, within the harmonic
approximation, produce an envelope function of oscillation given by \cite{phase}
\begin{eqnarray}
A = \bar{A}+\delta\!{A} & = & \left\{\begin{array}{ccc}
0 & , & t\le0 \\
\bar{A}\left(1-e^{-\beta t}\cos{\Omega_Rt}\right) & , & t>0
\end{array} \right. \; . \label{eq:Eq_25}
\end{eqnarray}
The two other modulated variables, $\delta\varphi_0=\gamma\delta\!{A}$
and $\bar{A}\delta\theta=\kappa\delta\!{A}$,
can then be found from the coefficients
\begin{eqnarray}
\gamma & = & \frac{-J_1(\bar{A})\sin\bar\varphi_0}
{\Omega^2-i\alpha\Omega-J_0(\bar{A})\cos\bar\varphi_0} \label{eq:Eq_26} \\
\kappa & = & \frac{\alpha\omega_s+2i\omega_s\Omega}
{\Omega^2-i\alpha\Omega+\omega_s^2-\omega_r^2} \; .
\label{eq:Eq_27}
\end{eqnarray}
The modulated system energy (for $\varepsilon_p(t)=0$) can be calculated from
Equation~(\ref{eq:Eq_28}) as
\begin{eqnarray}
\langle H\rangle_{\frac{2\pi}{\omega_s}} & = & \bar{H}+\delta{H} \; ,
\label{eq:Eq_34}
\end{eqnarray}
where $\bar{H}$ is the steady state energy of the phase-locked state in
Equations~(\ref{eq:Eq_10}) and (\ref{eq:Eq_11}), and $\delta{H}$ is the
transient modulation.
The average phase $\varphi_0$ can be calculated either as a function of
$A=\bar{A}+\delta\!{A}$ as given by Equations~(\ref{eq:Eq_8}) and
(\ref{eq:Eq_25}), or it can be extracted from the theory through
Equation~(\ref{eq:Eq_26}). The two results are very similar, but since the
latter is a first order approximation in the perturbation theory, it will
have a second order error at $t=0$. Using Equation~(\ref{eq:Eq_8}),
$\varphi_0(t)$ will satisfy the true value both at $t=0$ and for
$t\rightarrow\infty$.

\section{Simulation of Transients}
\begin{figure}[h]
\begin{center}
\scalebox{.30}{\centering \rotatebox{-90}{\includegraphics{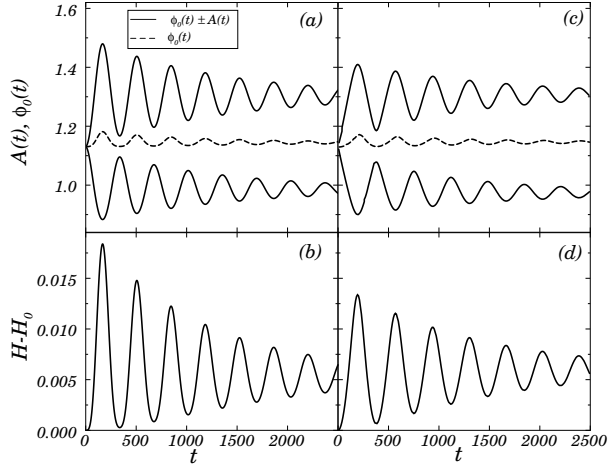}}}
\end{center}
\vspace{-0.0cm}
\caption{Normalized time response in phase amplitude $A$ and average
phase $\varphi_0$
(a and c) and normalized energy $H-H_0$ (b and d) as a function of time after
onset of microwave perturbation. Figures~a and b show results of the
perturbation theory Equations~(\ref{eq:Eq_8})-(\ref{eq:Eq_10}) and
(\ref{eq:Eq_25}). Figures~c and d show results
of direct simulations
of Equation~(\ref{eq:Eq_1}) for the same parameters as (a and b). Parameters
are: $\alpha=0.00151477$, $\eta=\sqrt{1-\omega_s^4}=0.904706$,
$\varepsilon_s(t)=0.00108\cdot\Theta(t)$, and $T=\varepsilon_p(t)=0$.
}
\label{fig:fig3}
\end{figure}

\begin{figure}[h]
\begin{center}
\scalebox{.30}{\centering \rotatebox{0}{\includegraphics{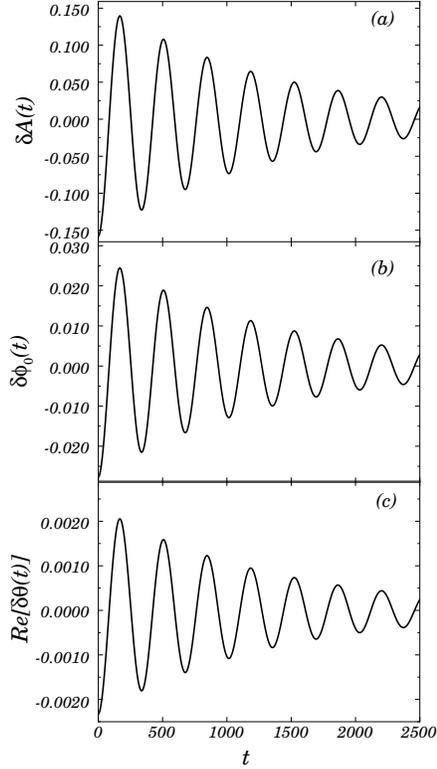}}}
\end{center}
\vspace{-1.0cm}
\caption{Linear transient responses to microwave onset versus normalized
time as calculated from
Equations~(\ref{eq:Eq_25})-(\ref{eq:Eq_27}) for the parameter values of
Figure~\ref{fig:fig3}.
}
\label{fig:fig4}
\end{figure}

\begin{figure}[h]
\begin{center}
\scalebox{.42}{\centering \rotatebox{0}{\includegraphics{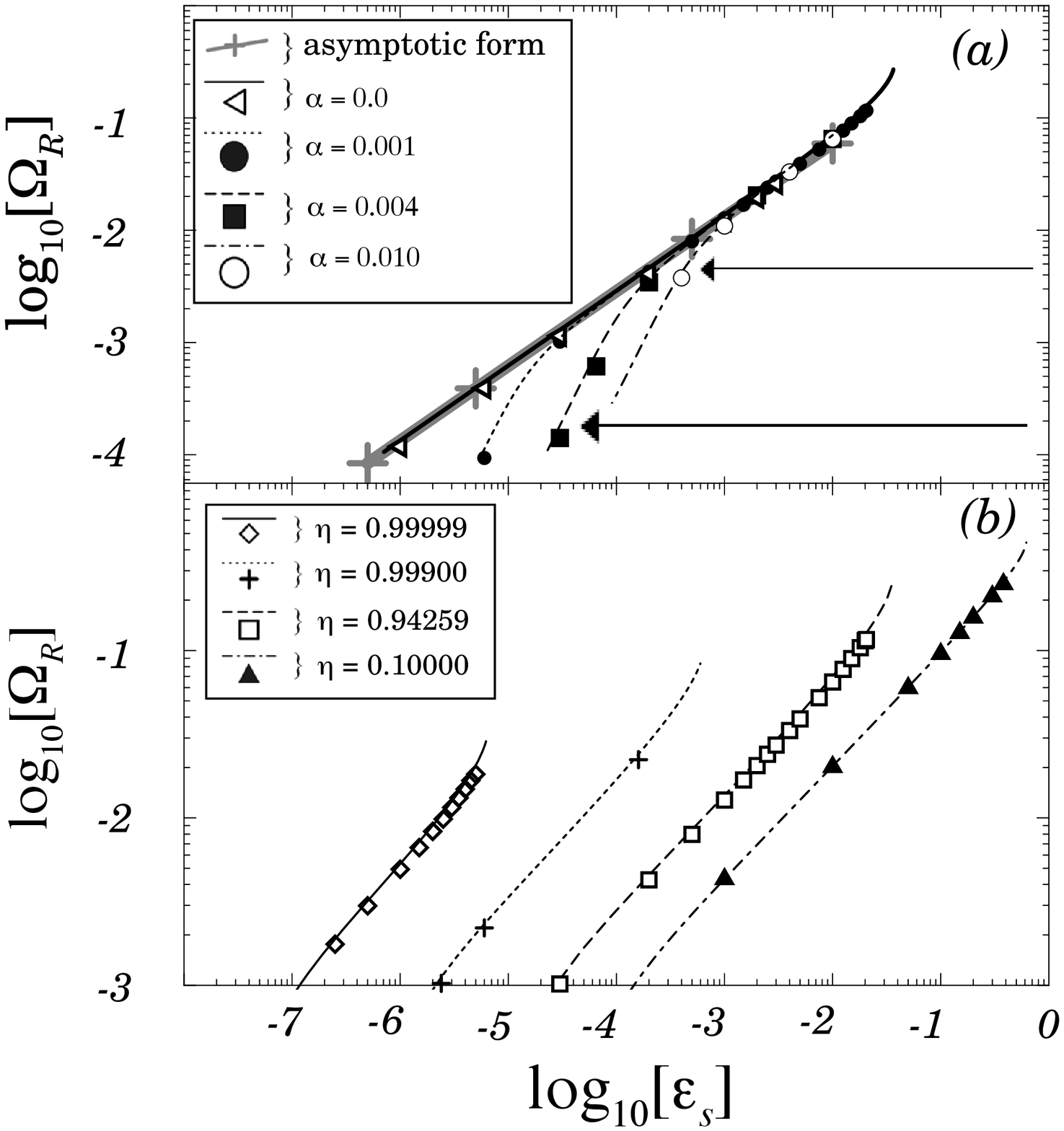}}}
\end{center}
\vspace{-3.0cm}
\caption{
Normalized modulation frequency $\Omega_R$ (at $T=0$) as a function of
normalized signal amplitude $\varepsilon_s$ for (a) a range of damping
   $\alpha$ values, with $\eta=0.94259$
($\omega_s = \sqrt[4]{1-\eta^2}=0.577886$). Gray line with large "+"
markers indicates the asymptotic form given in Equation~(\ref{eq:Eq_24}).
Arrows indicate the frequency for which small oscillations would be
overdamped for respective values of $\alpha$. (b) A range of dc bias
$\eta$ values, with $\alpha = 0.001$. Curves represent solutions to
Equation~(\ref{eq:Eq_16}). Markers represent data from numerical simulations
of Equation~(\ref{eq:Eq_1}) with $\varepsilon_p(t)=T=0$.
}
\label{fig:fig5}
\end{figure}

\begin{figure}[h]
\begin{center}
\scalebox{.42}{\centering \rotatebox{0}{\includegraphics{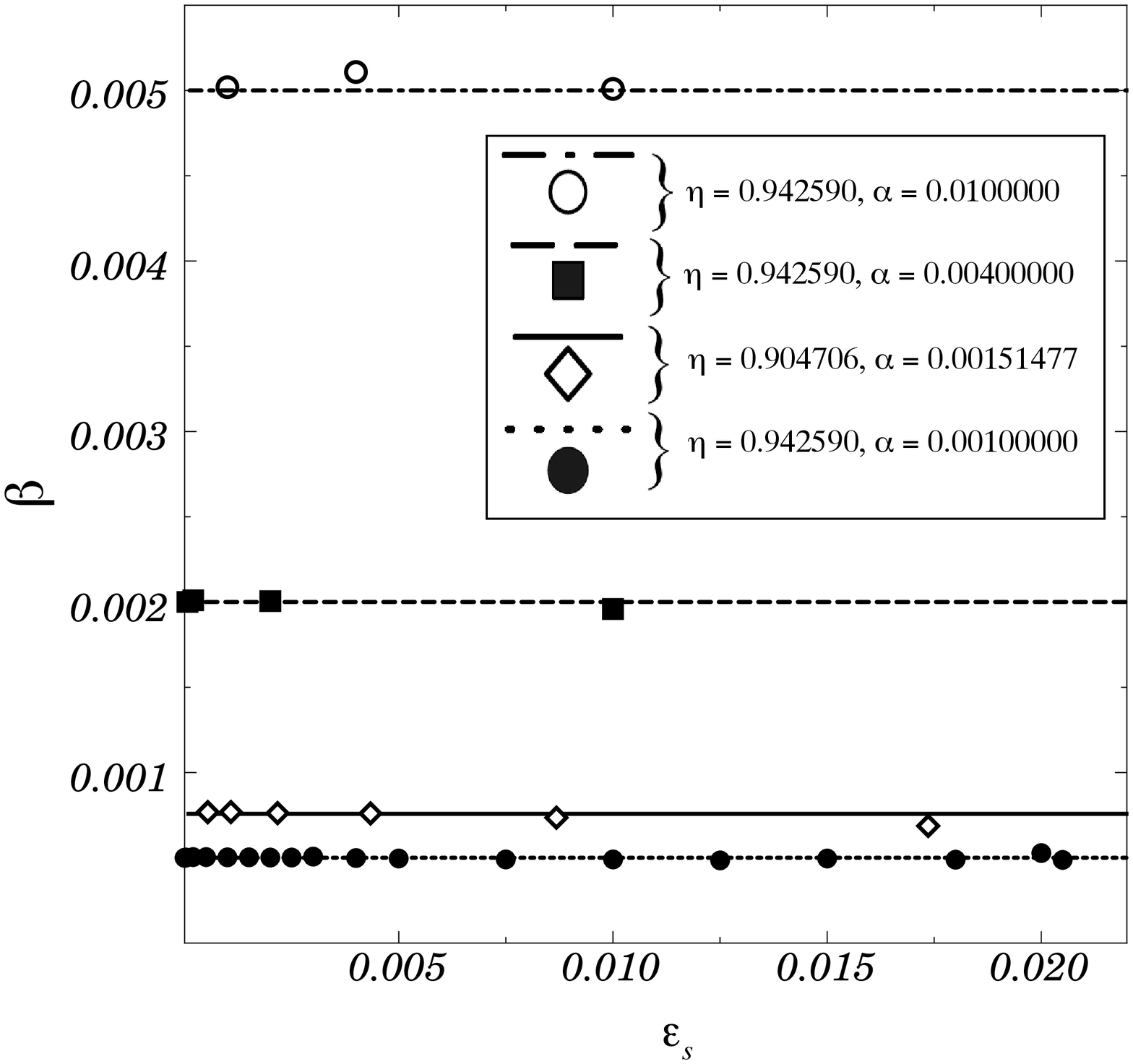}}}
\end{center}
\vspace{-3.0cm}
\caption{
Normalized attenuation $\beta$ as a function of damping
   $\alpha$ and normalized signal amplitude $\varepsilon_s$ for $T=0$. Lines represent
   the derived relationship $\beta = \alpha/2$ and markers represent data from
   the simulations of Figure~\ref{fig:fig5}.
}
\label{fig:fig6}
\end{figure}

Numerical validation of the expressions for the transient modulation can
be directly acquired from simulating Equation~(\ref{eq:Eq_1}) for
$\varepsilon_p(t)=T=0$ and $\varepsilon_s(t)=\varepsilon_s\Theta(t)$ for
different values of $\omega_s^4=1-\eta^2$ and $\alpha$. Simulations are
conducted by choosing $\eta<1$ and $\alpha\ge0$, then initiating the system in
the static state $(\varphi(0),\dot\varphi(0))=(\sin^{-1}\!\eta,0)$. At
$t=0$, the microwave field switches on, and we measure the phase $\varphi(t)$i,
the average phase
$\varphi_0(t)=\langle\varphi(t)\rangle_{\frac{2\pi}{\omega_s}}$, and the
system energy $H(t)$ as given by Equation~(\ref{eq:Eq_4}). Figure~\ref{fig:fig3}
shows a typical comparison between the analysis of the previous section
and direct numerical simulations. We have chosen system parameters
$\alpha=0.00151477$, $\eta=0.904706$, $\varepsilon_s=0.00108\cdot\Theta(t)$,
and $\omega_s=\sqrt[4]{1-\eta^2}=0.6527147$ (inspired by
Ref.~\cite{Claudon04}). We can clearly see that the transient
in phase and amplitude is a very slow modulation, and that this modulation
provides a slow transient oscillation in the system energy. Notice that the
phase-locked frequency is {\it much} higher than the depicted modulation,
$\Omega_R\ll\omega_s$. We have shown results of the theory on the left
(Figures~\ref{fig:fig3}a and b) and the corresponding simulation results
on the right (Figures~\ref{fig:fig3}c and d). The quantitative
agreement between the results is obvious. We do see a slightly larger
predicted oscillation amplitude compared to the simulated.
We also notice
a slightly higher predicted modulation frequency than what is observed
in the simulations. Despite these slight discrepancies, which are usually
about 5\% and no larger than about 10\%, we consider this remarkable
agreement given the simple monochromatic trial function and the
linearization in the perturbation treatment. Figure~\ref{fig:fig4}
shows the details
of the perturbation variables as calculated from the theory, and they reveal
that the actual oscillation frequency $\omega=\omega_s+\dot{\delta\theta}$
is only insignificantly different from $\omega_s$. Direct comparisons
between the predicted modulation $\Omega$ and the simulated values as a
function of microwave amplitude for different values of $\eta$ and
$\alpha$ are summarized in Figures~\ref{fig:fig5} and \ref{fig:fig6}.
Comparisons for $\Omega_R$ are provided in Figure~\ref{fig:fig5} and the
attenuation $\beta$ is validated in Figure~\ref{fig:fig6}.
The general agreement is very
good for all parameters tested. We observe the simple relationship
$\beta=\alpha/2$ for all oscillating solutions and the general trend of
the polynomial (Equation (\ref{eq:Eq_15}) or (\ref{eq:Eq_16}))
is to provide a modulation frequency slightly larger than
what is observed in simulations. This is quantitatively similar to the
values obtained from the energy balance perturbation theory developed in
Ref.~\cite{Jensen05}, where the true modulation frequency was consistently
slightly under-estimated. We finally observe the broad applicability of
the asymptotic value of the modulation frequency given in
Equation~(\ref{eq:Eq_24}).

\section{Rabi-Type Oscillations}
As was argued in Ref.~\cite{Jensen05}, the Rabi-oscillations observed in,
e.g., Refs.~\cite{Martinis02,Claudon04,Simmonds04} may be closely
related to the classical transient modulations described above. The
numerical simulations in Ref.~\cite{Jensen05} were conducted in close
agreement with the procedures described in Ref.~\cite{Martinis02}, by
applying a small temperature $T\approx50mK$ ($kT/H_J\sim10^{-2}\ll1$)
to a Josephson junction,
which is perturbed by a microwave pulse starting at $t=0$ with a frequency
$\omega_s=\sqrt[4]{1-\eta^2}$. Varying either the microwave amplitude or
duration may yield Rabi-type oscillations in switching probability
when a subsequent microwave pulse with a smaller frequency
$\omega_p<\omega_s$ is applied to probe the energy state of the junction
as it is left by the first microwave pulse. The frequency of the second
(probe) field was chosen to match an anharmonic amplitude that could
excite the junction beyond the energy saddle point and lead to escape
from the well.
The measurements were conducted for varying microwave amplitude and fixed
microwave duration.

\begin{figure}[h]
\begin{center}
\scalebox{.42}{\centering \rotatebox{0}{\includegraphics{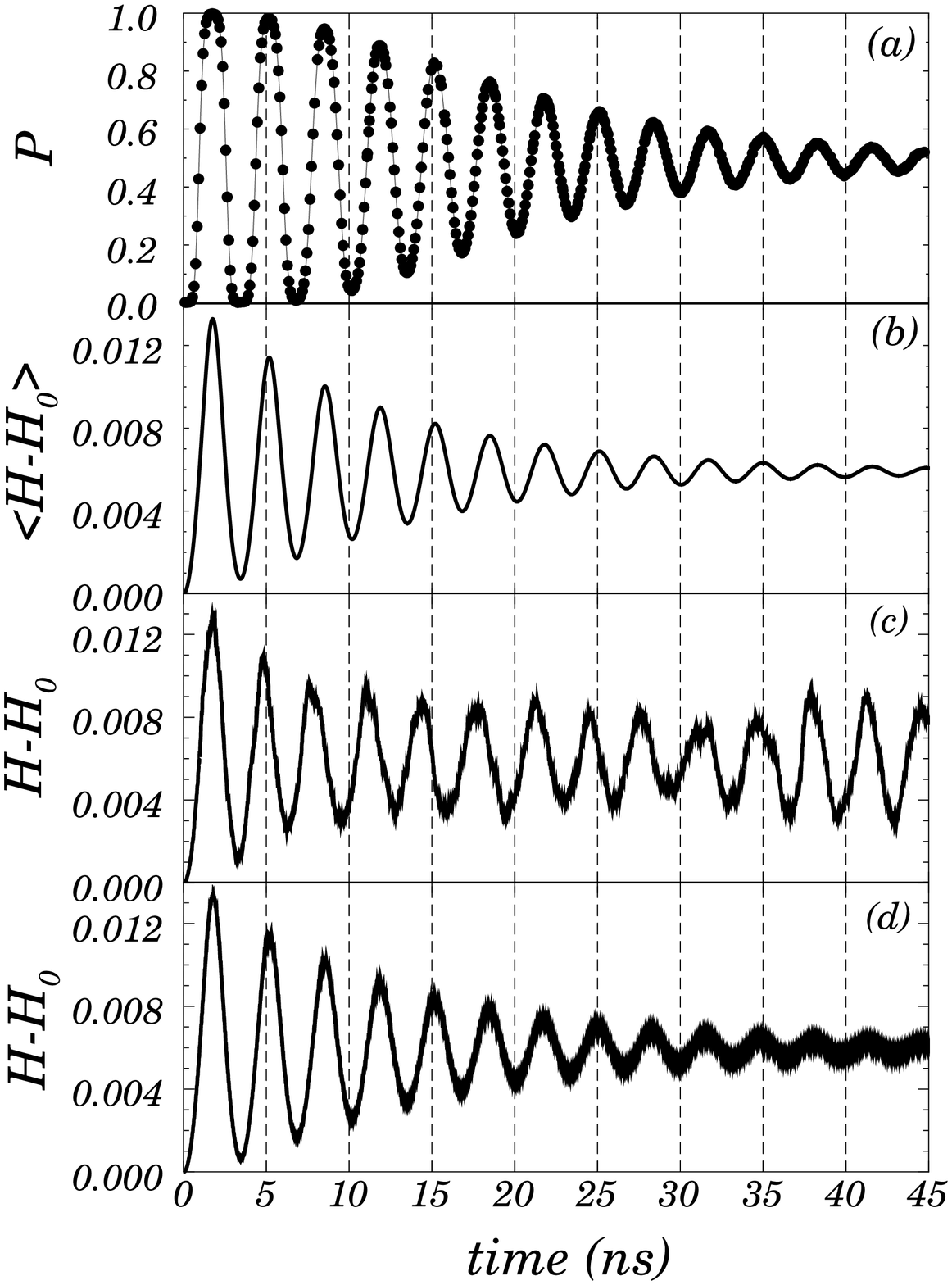}}}
\end{center}
\vspace{-1.0cm}
\caption{
(a) Escape probability $P$ as a function of signal duration
   $\Delta t_s\omega_0^{-1}$. Numerically simulated quantities are represented
   by dots (each of which represents 25,000 escape events).
(b) Ensemble average of normalized energy, simulated through
Equation~(\ref{eq:Eq_1}), as a
   function of signal duration (ensemble size N = 50,000, randomly chosen
$\theta_s$). Parameters are:
$T = 30$ mK, $\alpha=0.00151477$, $\eta=\sqrt{1-\omega_s^4}=0.904706$,
   $\omega_s=2\pi\nu_{01}/\omega_0=0.652714$, $\varepsilon_s=0.00217000$,
   $\varepsilon_p=0.0847400$, and $\omega_0\approx110\cdot10^9s^{-1}$.
These parameters are inspired by those
reported in Ref.~\cite{Claudon04}. (c) Single trajectory of normalized
energy versus time for parameters listed for (b).
(d) Single trajectory of normalized energy versus time for $T=0$.
}
\label{fig:fig7}
\end{figure}

\begin{figure}[h]
\begin{center}
\scalebox{.42}{\centering
\rotatebox{0}{\includegraphics{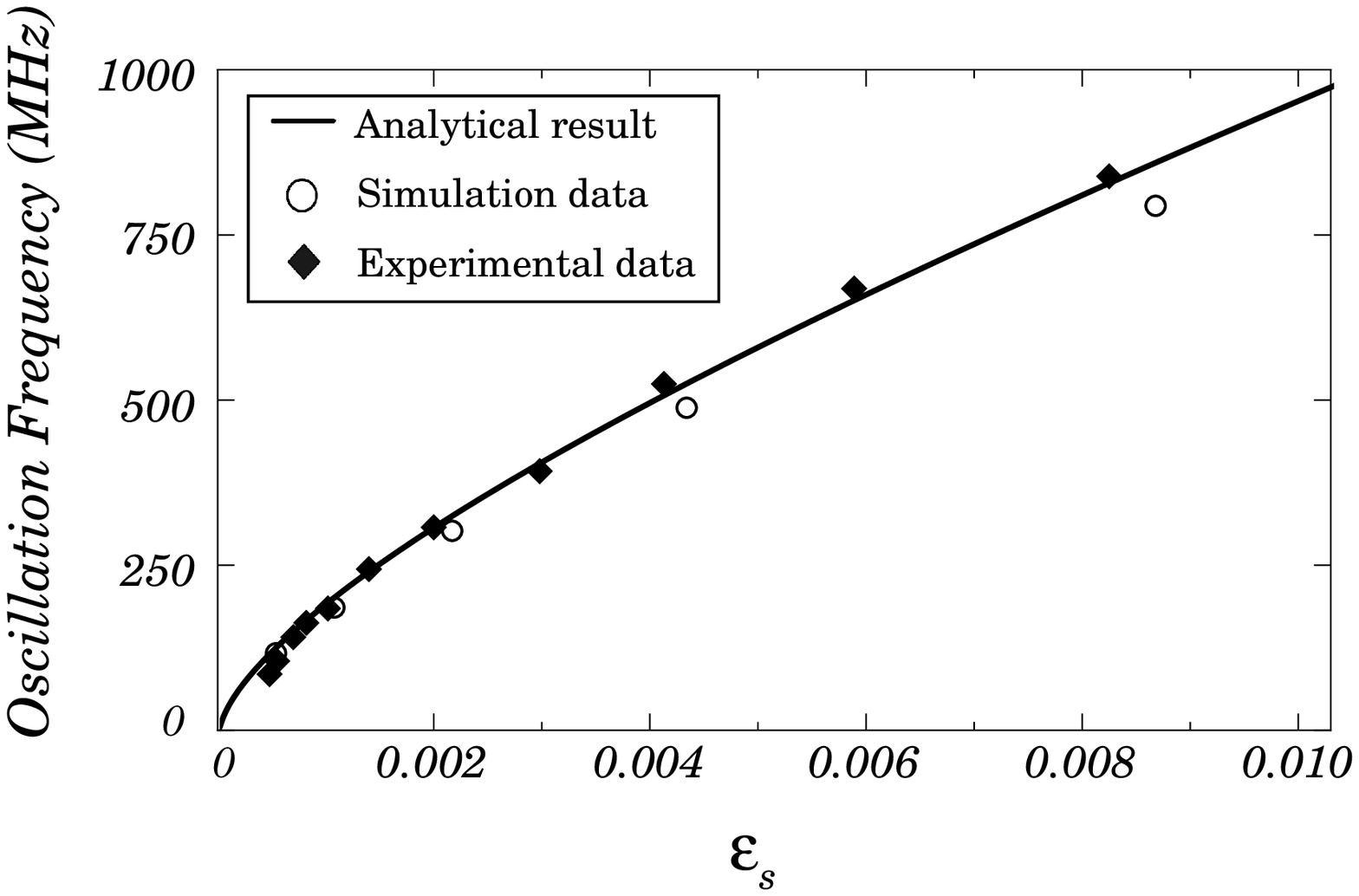}}}
\end{center}
\vspace{-7.0cm}
\caption{
Modulation frequency $\Omega_R$ as a function of normalized signal amplitude
$\varepsilon_s$ for $\alpha = 0.00151477$, $\eta = 0.904706$, $T=30$ mK.
Lines represent calculations using Equation~(\ref{eq:Eq_16}), $\circ$ represent
statistical data from simulations, and $\bullet$ are the measurements copied
from Ref.~\cite{Claudon04}.
}
\label{fig:fig8}
\end{figure}

The experiments described in Ref.~\cite{Claudon04} were conducted slightly
differently. First, the Josephson system was a small-inductance $\beta_L$
interferometer, which we can conveniently approximate with a single Josephson
junction, since the dynamics for small $\beta_L$ is known to be well
represented by a single
degree of freedom \cite{Jensen03}. Second, the probe pulse was in this case
a "dc" pulse as sketched in Figure~\ref{fig:fig1}, and the measurements
were presented for
fixed microwave amplitude and variable duration. We will here provide
simulations of the classical system described in Equation~(\ref{eq:Eq_1}) and
parameterized by information provided in Ref.~\cite{Claudon04}.
Characteristic current and frequency are $I_c=6.056\mu{A}$ and
$\omega_0\approx110\cdot10^{9}s^{-1}$, which
lead to a normalized temperature of $kT/H_J\approx2\cdot10^{-4}$ ($T=30mK$),
and a normalized microwave frequency $\omega_s=0.6527147$, which is
close to the resonance such that $\eta=\sqrt{1-\omega_s^2}$. We have
estimated the dissipation parameter $\alpha$ from the reported decay
of Rabi-oscillations, $\alpha=0.00151477$.
The normalized applied microwave amplitude
$\varepsilon_s$ is varied between 0 and $0.01$, and the normalized duration
is in the range $[0;3000]$. Simulations are conducted with a probe pulse
$\varepsilon_p(t)$ as shown in Figure~\ref{fig:fig1}, followed by a short 
time in which we determine whether or not the junction has switched from
the zero-voltage state. For every specific set of parameters,
this type of simulation is conducted 25,000 times, each with a randomly chosen
value of microwave phase $\theta_s$, and the switching probability is recorded,
before a new microwave duration is chosen. Typical results are displayed in
Figure~\ref{fig:fig7}a,
where the switching probability $P$, reported as population occupancy of excited
quantum state in Ref.~\cite{Claudon04}, is shown as a function of microwave
duration. We clearly observe the Rabi-type oscillations of this classical
system with a distinct frequency. Moreover, the oscillation frequency
is in near perfect agreement with the reported comparable figure of the
experimental measurements as well as the
theoretical value $\Omega_R$ provided by the analysis in the above section.
Figure~\ref{fig:fig7}b shows the simulated system energy as an average
of many thermal realizations, each with a randomly chosen $\theta_s$,
of the trajectory (at $T\approx30mK$). The different choices
of $\theta_s\in[-\pi;\pi]$ yield slightly different trajectories of
$\delta\!{A}(t)$, $\delta\theta(t)$, and $\delta\varphi_0(t)$, which in turn
make possible switching a function of $\theta_s$. The
importance of this ensemble average can be viewed in Figure~\ref{fig:fig7}c,
where a single trajectory of energy is shown for the applied temperature
of $T=30mK$. After a short transient, the thermal effects are dominating
the individual trajectory, exciting the modulation frequency $\Omega_R$ at
random phase. Only after an appropriate average over trajectories do we
observe the smoothly decaying envelope of the modulated
energy curve in Figure~\ref{fig:fig7}b, which
in turn is very similar to the (single) zero-temperature trajectory seen in
Figure~\ref{fig:fig7}d. The
close connection between the transient energy modulation and the Rabi-type
modulation is obvious, and the close relationship between classical numerical
simulations, classical theory,
and the experimental measurements is further emphasized in
Figure~\ref{fig:fig8}, where
the experimental data (open markers) of Ref.~\cite{Claudon04} are
shown alongside theory (solid curve) and simulations (closed markers).
The three-way close agreement indicates that much about these measurements
can be understood from classical theory of driven, damped Josephson junctions.

\section{Conclusion}
We have presented a perturbation analysis of the classical,
nonlinear model describing the microwave-perturbed Josephson junction.
Results show direct quantitative analogy between experimentally reported
Rabi-oscillations and the classical transients to phase-locking, and
results are presented as specific functions of experimental system parameters.
The analysis in this paper is both a complement and an extension to the
analysis of Ref.~\cite{Jensen05}, where an energy balance perturbation approach
was applied to produce similar connection between reports of Rabi-oscillations
and the nonlinear classical Josephson system. The extension provides
direct closed form solutions of a (Rabi-type) modulation frequency as well
as attenuation, and the resulting frequency agrees closely with the previously
obtained result. This consistency lends credibility to the value of
a classical interpretation of the reported Rabi-oscillations, since the
quantum mechanical and classical intuition turns out to be very similar
(as is also the case in laser physics \cite{Zhu90}).

The macroscopic quantum picture interprets the observed oscillations
as a result of a microwave induced temporal variation
in the population probability of two, or more, intrinsic quantum mechanical
energy levels in the Josephson washboard potential. This slow variation is
then indirectly observed through the associated variation in tunneling
probability, switching from the zero-voltage state as a result of the
application of the probe pulse. The stochastic nature of the system is
due to the absorption of microwave photons as well as the
subsequent tunneling probability of the measurement. 

The classical picture
presents a system with a microwave induced temporally modulated energy. 
The slow variation is indirectly observed through the associated variation in
escape from the energy well when the probe is applied, with high
probability for passing the energetic saddle point during times of
elevated energy and relatively little escape probability during times
of depressed energy content. The stochastic nature of this system is
due to the random phase of the microwave signal as well as the thermal
fluctuations. 

Additional analogies between intrinsic quantum mechanical
energy levels and the multi-valued resonances of the classical nonlinear
system briefly outlined above (as well as in Ref.~\cite{Jensen05}) further
bridges the gap between what can be expected from the two
interpretations of the microwave induced measurements. Since the
experimental measurements are concerned only with detecting the escape
event, and since the classical and quantum mechanical interpretations
seem to provide the same signatures for that escape, we are faced with a
fundamental ambiguity of how to read this information.
Potential applications of Josephson technology for quantum information
processing will therefore benefit from the development of new unambiguous
measurements, which must present signatures of macroscopic quantum
behavior that cannot be explained classically. However, we submit that,
given the very close quantitative agreement between our classical theory
and the experimental measurements, the intuition and the closed form
expressions presented in this paper can be directly applied to guide
future experiments and microwave manipulations of Josephson systems.

\section{Acknowledgments}
This work was supported in part by the UC~Davis Center for Digital Security
under AFOSR grant FA9550-04-1-0171 and in part by MIUR (Italy) COFIN04.


\begin{thebibliography}{99}
\bibitem{phase} This simple expression is not sensitive to the microwave
phase $\theta_s$, which does have some influence on the transient oscillations
(to be published).
\end{thebibliography}
\end{document}